\documentclass[preprint,aip,amssymb,amsmath,nobibnotes,floatfix,12pt]{revtex4-1}
\usepackage{bm}%
\usepackage{etex}
\usepackage[all]{xy}
\usepackage[T1]{fontenc}
\usepackage[utf8]{inputenc}
\usepackage[english]{babel}
\usepackage[english]{varioref}
\usepackage{amsthm}
\usepackage{amsmath}
\usepackage[version=4]{mhchem}
\usepackage{graphics} 
\usepackage{subfig}
\usepackage{stmaryrd}
\usepackage{amssymb}
\usepackage{amscd}
\usepackage{tensor}
\usepackage{hyperref}
\usepackage{mathtools}
\DeclareMathAlphabet{\mathpzc}{OT1}{pzc}{m}{it}
\usepackage{mathrsfs}
\newtheorem{theorem}{Theorem}
\DeclareMathOperator{\diag}{diag}
\newcommand{\gv}[1]{\ensuremath{\mbox{\boldmath$ #1 $}}}
\newcommand{\op}[1]{\hat{\mathscr{#1}}} % for operators
\newcommand{\sop}[1]{\hat{\hat{\mathfrak{#1}}}} % for superoperators
\newcommand{\ket}[1]{\left| #1 \right>} % for Dirac ket
\newcommand{\rket}[1]{\left| #1 \right)} % for Dirac-type reducible kets
\newcommand{\matrixel}[3]{\left< #1 \vphantom{#2#3} \right|
 #2 \left| #3 \vphantom{#1#2} \right>} % for Dirac matrix elements
\newtheorem{corollary}{Corollary}[theorem]
\newtheorem{prop}{Proposition}
\usepackage[sort&compress]{natbib}
\begin{document}
\preprint{IOP/123-QED}

\title{Magnetic Resonance, Index Compression Maps and the Holstein-Primakoff Bosons:\\ Towards a Polynomially Scaling Exact Diagonalization of Isotropic Multispin Hamiltonians.}

\author{J.A. Gyamfi\footnote{Email corresponding author: \href{mailto:jerryman.gyamfi@sns.it}{jerryman.gyamfi@sns.it} }}
\affiliation{Scuola Normale Superiore di Pisa, Piazza dei Cavalieri 7, 56126 Pisa, Italy.}

\author{V. Barone}
\affiliation{Scuola Normale Superiore di Pisa, Piazza dei Cavalieri 7, 56126 Pisa, Italy.}

\date{\today}

\begin{abstract}
Matrix diagonalization has long been a setback in the numerical simulation of the magnetic resonance spectra of multispin systems since the dimension of the Hilbert space of such systems grows exponentially with the number of spins -- a problem commonly referred to as the "curse of dimensionality". In this paper, we propose two mathematical instruments which, when harmoniously combined, could greatly help surmount to a fair degree and in a systematic manner the curse of dimensionality. These are: 1) the Holstein-Primakoff bosons and 2) what we have termed the "index compression maps". These two allow a bijective mapping of (multi)spin states to integers. Their combination leads to the block diagonalization of the multispin Hamiltonian, thus a computationally exact way of diagonalizing the latter but which also reduces significantly the computational cost. We also show that the eigenvectors and eigenvalues of the Liouvillian operator can be easily determined once those of the related multispin Hamiltonian are known. Interestingly, the method also enables an analytical characterization of the multispin Hilbert space -- a feat hardly attainable with other approaches.
\par We illustrate the method here by showing how a general static isotropic multispin Hamiltonian could be exactly diagonalized with very less computational cost. Nonetheless, we emphasize that the method could be applied to study numerous quantum systems defined on finite Hilbert spaces and embodied with at most pairwise interactions.
\end{abstract}

\pacs{}% insert suggested PACS numbers in braces on next line

\maketitle %\maketitle must follow title, authors, abstract and \pacs

\section{Introduction}
It is of common knowledge that the computational cost of simulating many-body quantum systems on classical computers grow exponentially with the dimension $D$ of the system's Hilbert space, making the computation hardly feasible when $D$ exceeds some threshold\citep{art:Feynman-1982}. Such computational headaches fall under what has been christened the "curse of dimensionality".
\par When it comes to simulating the magnetic resonance spectrum (say EPR or NMR) of a chemical species, the problem is particularly felt and somehow frustrating because, on top of the exponential growth of the computational cost, one ends up with even more sparse matrices as the number of nonzero spins in the system increases. 
\par The relentless growth of the sparseness of the matrices is an indication that the dynamics of these systems are effectively encoded in certain subspaces of the Hilbert space, whose dimensions in most cases scale polynomially with the number of particles. Thus, overcoming the curse of dimensionality is, in the long run, equivalent to finding an efficient way of determining the said subspaces and restricting to them the numerical simulation of the system's dynamics. In magnetic resonance, computational protocols like the adaptive state-space restriction (SSR) method proposed by Kuprov \emph{et. al.}\cite{art:Kuprov-2007}, for example, constructs an effective subspace by excluding entanglements beyond a certain fixed order, thus yielding an effective subspace which scales polynomially in the total number of spins $N$. We show here that it is possible to devise a computationally exact diagonalization scheme for static isotropic multispin Hamiltonians which reduces significantly the computational cost, without excluding any states or interactions. The method hints at the possibility of a polynomially scaling numerical protocol for the simulation of multispin dynamics, and yet more analytical in nature. 
\par We are of the view that the curse of dimensionality can be fairly dealt with if one could eliminate superfluous mathematical operations during numerical simulations. But this requires adequate analytical tools. We may say the mathematical operation at the core of the reason why the computational cost of multispin numerical simulations grows exponentially with the number of spins is the Kronecker product. It would thus be most desirable if we have a set of analytical instruments which could help us "navigate" confidently through this mathematical maze. It should therefore not be of any surprise if central to the method we present here are: 1) the Holstein-Primakoff bosons, and 2) what we have termed the "row index compression map" $\eta$. Together, they expedite the determination of new good quantum numbers -- besides allowing an easy monitoring of the various spin transitions in the system. In addition, $\eta$ releases us from the burden of carrying along long strings of indexes during theoretical derivations, making them manageable -- not to mention its usefulness in numerical simulations. The great benefit of the index compression maps we discuss below is that they enable us to construct the Kronecker product at the local level, which presents great advantages. This will be clear very soon. 
\par Though we focus here on multispin systems and magnetic resonance, the method could be easily employed in other problems, provided the Hilbert space is finite and presents at most pairwise interactions.
\par The paper is organized as follows: After a brief review of the notations we have adopted (\S \ref{sec:Notations}), we first consider the index compression map theorem and its consequences in \S \ref{sec:Index_compression_maps}. This is followed by a brief review of the Holstein-Primakoff transformation in \S \ref{sec:HP_transformation}. With these two instruments in hand, we illustrate how they could be used to efficiently diagonalize in a computationally exact manner static isotropic multispin Hamiltonians with less computational cost compared to traditional methods in \S \ref{sec:Row_HP_diag}. In \S \ref{sec:Application}, we apply our results to briefly discuss the diagonalization of the isotropic static multispin Hamiltonian of the deuterated hydroxymethyl radical (\ce{^{.}CH2OD}).

\section{Notations and terminology}\label{sec:Notations}
\par Consider a multiset $\mathpzc{A}$ of $N$ spins: $\mathpzc{A}:=\{j_1, j_2, \ldots , j_N\} \equiv \{j^{N_\alpha}_\alpha\}$, where $\alpha \in \{ 1, 2, \ldots , \sigma\}$. The index $\alpha$ certainly runs over the \emph{distinct} elements (in total $\sigma$) of $\mathpzc{A}$, and $N_\alpha$ is the multiplicity of $j_\alpha$. Notably, $N = \sum^\sigma_{\alpha = 1} N_\alpha$. By distinct elements we mean $SU(2)$ spins with different maximal weights -- other characteristics of the spins (like charge, magnetic moment etc.) are not of any merit whatsoever here. For example, the multiset $\mathpzc{A}=\{\frac{1}{2}^9, 1^3\}$ indicates any aggregate of nine spin-$1/2$ and three spin-$1$; the actual composition could consist of, for instance, 1) nine electrons and three protons, or 2) five \ce{^{15}_{7}N_8}, four muons -- both spin-$1/2$s -- and three \ce{^{14}_{7}N_7} (spin-$1$)\cite{art:Fuller-1976}, etc. In the following, we reserve the Latin letter $i$ to lower index $\mathpzc{A}$'s elements when the latter is understood as $\mathpzc{A}=\{j_1, j_2, \ldots , j_N\}$ , while we use the Greek index $\alpha$ when we mean distinct elements of $\mathpzc{A}$.
\par Operators will be indicated with their usual hats while their matrix representations will bear none: for example, the matrix representation of the operator $\op{A}$ will be simply indicated as $\mathscr{A}$.
\par Finally, we shall refer to nonzero submatrices as simply "submatrices".
\section{The index compression maps, $\eta$ and $\xi$.}\label{sec:Index_compression_maps}
The Hilbert space $\mathcal{H}$ of a given multispin system $\mathpzc{A}$ is constructed by taking the direct product of the Hilbert space $\mathcal{H}_i$ of each individual spins. The basis (in the uncoupled representation) of $\mathcal{H}$ is determined in similar fashion. The index compression maps (in particular, the row index one) we explore below allow one to determine  the composition of each basis element of the global Hilbert space in terms of the basis elements of the separate Hilbert spaces $\mathcal{H}_i$. In addition, we gain a bijective mapping between $\mathcal{H}$'s basis and the set of natural numbers. This is particularly useful as we shall see in the next section.
\par We present below the index compression map theorem in general terms. Its application and usefulness go far beyond the applications we propose in this paper. 
\begin{theorem}[The index compression map theorem]\label{thm:basis-numbers-mapping}
Given the finite set of finite matrices $\{A_1, A_2, \ldots , A_N\}$ -- where the dimension of the matrix $A_i$ is $m_i \times n_i$, $\left(m_i , n_i \in \mathbb{N}\right)$ -- and the ordered Kronecker product
					\begin{subequations} 
					\begin{align}
					C & = A_1 \otimes  \cdots \otimes A_{N-1} \otimes  A_N \label{eq:matrix_C}\\ 
					c_{\nu , \mu}  & = a_{{\nu}_1 ,\hspace{0.03cm} {\mu}_1}\ \cdots \ a_{{\nu}_{N-1},\hspace{0.03cm} {\mu}_{N-1}}  \ a_{{\nu}_N, \hspace{0.03cm} {\mu}_N} \label{eq:c_nu_mu}\ ,
					\end{align}
					\end{subequations}		
then, there is a one-to-one correspondence between the index $\nu$ ($\mu$) and the indexes $\{\nu_1, \ldots ,\nu_{N-1}, \nu_{N} \}$ ($\{ \mu_1, \ldots ,\mu_{N-1}, \mu_{N}\}$) given by the mapping $\eta$ ($\xi$), such that
					\begin{equation}
					\label{eq:nu_row}
					\begin{split}
					\eta & : \mathbb{G}_r \left( \subset \underbrace{\mathbb{N}_0 \times \mathbb{N}_0 \times \cdots \times \mathbb{N}_0}_{N}\right) \longrightarrow  \mathbb{N}_0  \\
					\nu & = \eta (\nu_1 , \ldots , \nu_N) = 1 + \sum^N_{i=1} \mathpzc{W}_{R,i} (\nu_i - 1)
					\end{split}
					\end{equation}
					
					\begin{equation}
					\label{eq:mu_column}
					\begin{split}
					\xi & : \mathbb{G}_c \left( \subset \underbrace{\mathbb{N}_0 \times \mathbb{N}_0 \times \cdots \times \mathbb{N}_0}_{N} \right) \longrightarrow  \mathbb{N}_0 \\
					\mu & = \xi (\mu_1 , \ldots , \mu_N) =  1 + \sum^N_{i=1} \mathpzc{W}_{C,i} (\mu_i - 1) \ 
					\end{split}
					\end{equation}
where,
					\begin{subequations}
					\label{eq:def_W}
					\begin{align}
					\mathpzc{W}_{R,i}  & \equiv \delta_{N,i} + (1-\delta_{N,i})\prod^{N-1-i}_{l=0} m_{N-l} \label{eq:def_W_R}\\
					\mathpzc{W}_{C,i}  & \equiv \delta_{N,i} + (1-\delta_{N,i})\prod^{N-1-i}_{l=0} n_{N-l}  \ .
					\end{align}
					\end{subequations}
\end{theorem}
The entries of the matrix $C$ resulting from the Kronecker product given in Eq. \eqref{eq:matrix_C} are products which have as factors an entry from each factor matrix on the RHS.  The mapping $\xi$, termed the "column index compression map", assigns a unique column integer to the (ordered) collection of column indexes of the factors. Similarly, $\eta$ maps a string of ordered row indexes of the factors to a unique row index of $C$. $\mathbb{G}_r$ ($\mathbb{G}_c$) is the set of all possible ordered combinations of row (column) indexes of the factor matrices. It can be proved that both $\eta$ and $\xi$ are bijective, thus with inverses. That is, given $\nu $, for example, one can find the corresponding string of ordered row indexes of the factor matrices.
\par Take for example the following Kronecker product:
				\begin{subequations}
				\label{eq:C_example}
				\begin{align}				
				C & = A \otimes B \otimes D \label{eq:C_example_a} \\
				c_{\nu ,\mu} & = a_{\nu_1 ,\mu_1} b_{\nu_2, \mu_2} d_{\nu_3, \mu_3} 
				\end{align}
				\end{subequations}
where,
				\begin{equation}
				A= \begin{pmatrix}
				a_{1,1} & a_{1,2} & a_{1,3} \\
				a_{2,1} & a_{2,2} & a_{2,3} \\
				a_{3,1} & a_{3,2} & a_{3,3} 
				\end{pmatrix}
					\qquad
				B= \begin{pmatrix}
				b_{1,1}\\
				b_{2,1}\\
				b_{3,1}
				\end{pmatrix}
					\qquad
				D= \begin{pmatrix}
				d_{1,1} & d_{1,2} \\
				d_{2,1} & d_{2,2}
				\end{pmatrix} \ .
				\end{equation}
We have here desisted from denoting the factor matrices as $A_1, A_2$ and $A_3$ in order to simplify the notations, but it is clear that the matrices $A, B, D$ in Eq. \eqref{eq:C_example_a} correspond to $A_1,A_2$ and $A_3$ in Eq. \eqref{eq:matrix_C}, respectively. The map $\eta$ relates any given set of row indexes $\nu_1, \nu_2,\nu_3$ of the matrices $A,B,D$, respectively, with a specific row index $\nu$ of $C$. The map $\xi$ does an analogous thing but with the column indexes $\mu_1, \mu_2, \mu_3$ and $\mu$. Now, from Eqs. \eqref{eq:nu_row} - \eqref{eq:def_W}, it follows that for the particular Kronecker product given in Eq. \eqref{eq:C_example_a}, the maps $\eta(\nu_1,\nu_2,\nu_3)$ and $\xi(\mu_1,\mu_2,\mu_3)$ are respectively of the form,
				\begin{subequations}
				\label{eq:nu_mu_examples}
				\begin{align}
				\nu & = \eta(\nu_1,\nu_2,\nu_3)= 1 + 6 (\nu_1-1) + 2(\nu_2-1 ) + (\nu_3-1) \ , \label{eq:nu_example}\\
				\mu & = \xi(\nu_1,\nu_2,\nu_3)= 1 + 2 (\mu_1-1) + 2(\mu_2-1 ) + (\mu_3-1) \label{eq:mu_example}\ .
				\end{align}
				\end{subequations}
Say we want to know which element of $C$ corresponds to the product $a_{2,1}b_{3,1}d_{2,2}$. In other words, we want to know $\nu$ and $\mu$ given that $\nu_1=2, \mu_1=1$ (since the element of $A$ contributing to the product is $a_{2,1}$), $\nu_2=3, \mu_2=1$ (given that we have $b_{3,1}$), and $\nu_3=2, \mu_3=2$. Inserting these integers into Eqs. \eqref{eq:nu_mu_examples}, we find that $\nu=\eta(2,3,2)=12$ and $\mu=\xi(1,1,2)=2$. Hence, $a_{2,1}b_{3,1}d_{2,2}=c_{12,2}$. 
\par In the other way round, we may have $\nu$ and $\mu$ of $c_{\nu, \mu}$ and may want to know which component of $A,B,D$ gives rise to that specific element of $C$. For example, we may ask what are the elements of $A,B,D$ whose product gives rise to the element $c_{9,5}$ of $C$. This corresponds  to solving the Diophantine equations
				\begin{subequations}
				\label{eq:nu_mu_examples_2}
				\begin{align}
				9 & = \eta(\nu_1,\nu_2,\nu_3)= 1 + 6 (\nu_1-1) + 2(\nu_2-1 ) + (\nu_3-1) \ , \label{eq:nu_example_2}\\
				5 & = \xi(\mu_1,\mu_2,\mu_3)= 1 + 2 (\mu_1-1) + 2(\mu_2-1 ) + (\mu_3-1) \label{eq:mu_example_2}\ .
				\end{align}
				\end{subequations}
for $\nu_1,\nu_2,\nu_3,\mu_1,\mu_2,\mu_3$. The dimensions of the factor matrices $A,B,D$ serve as constraints on the unknown variables. A quick strategy in solving these Diophantine equations is the following: Consider for example Eq. \eqref{eq:nu_example_2}. Find the integer among the possible values of $\nu_1$ which maximizes the sum of the first two terms; the sum must not be greater than $\nu$, in this case $9$. Find then the integer in the range of $\nu_2$ which maximizes the first three terms of Eq. \eqref{eq:nu_example_2}, with $\nu_1$ equal to the integer found in the previous step. Here too, the sum must not exceed the value of $\nu$. Continue the process until you get to $\nu_N$, which must be such that its value, together with the those of $\nu_1, \ldots, \nu_{N-1}$ found earlier, gives exactly $\nu$ according to Eq. \eqref{eq:nu_row}. The same strategy applies in the case of the column index $\mu$. As for Eq. \eqref{eq:nu_example_2}, we find that $\nu_1=2$, $\nu_2=2$ and $\nu_3=1$. Analogously, we find that for Eq. \eqref{eq:mu_example_2} to hold, the following must be true: $\mu_1=3, \mu_2=\mu_3=1$. Thus, $c_{9,5}=a_{2,3} b_{2,1} d_{1,1}$, which can be easily verified. For the procedure outlined to give the right solution, it is important that the summands of $\eta$ and $\xi$ are written in the same order as done on the RHS of Eqs. \eqref{eq:nu_example_2}-\eqref{eq:mu_example_2}, \emph{i.e.} the constant term must appear first, then the $\nu_1$ (or $\mu_1$) term, and so on. To understand why this is so, it is useful to realize that the compression maps $\eta$ and $\xi$ may be interpreted as functions which convert string of numbers from a "multi-numeral" base to the decimal base (for now, the integer zero is not an allowed digit). 
\par It is important to note that the order of the factor matrices determine the equation for the maps $\eta$ and $\xi$. Indeed, a permutation of the factor matrices effects a permutation of the elements of the resulting matrix. For example, if instead of the product in Eq. \eqref{eq:C_example_a} we consider the following product
				\begin{subequations}
				\label{eq:C_example_3}
				\begin{align}				
				C' & = B \otimes A \otimes D \label{eq:C_example_3}\\
				c'_{\nu' ,\mu'} & = b_{\nu'_1 ,\mu'_1} a_{\nu'_2, \mu'_2} d_{\nu'_3, \mu'_3}
				\end{align}
				\end{subequations}			
where with respect to  Eq. \eqref{eq:C_example_a} we have swapped the matrices $B$ and $A$, we note that the mappings $\eta$ and $\xi$ transform as
				\begin{subequations}
				\label{eq:nu_mu_examples_3}
				\begin{align}
				\nu' & = \eta(\nu'_1,\nu'_2,\nu'_3)= 1 + 6 (\nu'_1-1) + 2(\nu'_2-1 ) + (\nu'_3-1) \ , \label{eq:nu_example_3}\\
				\mu' & = \xi(\nu'_1,\nu'_2,\nu'_3)= 1 + 6 (\mu'_1-1) + 2(\mu'_2-1 ) + (\mu'_3-1) \label{eq:mu_example_3}\ .
				\end{align}
				\end{subequations}
With Eq. \eqref{eq:C_example_3}, the product $a_{2,1}b_{3,1}d_{2,2}=c'_{16,2}$. Nonetheless, $a_{2,3} b_{2,1} d_{1,1}$ is still $c'_{9,5}$.
\par As one can see, the lowest value of both $\nu$ and $\mu$  as defined in Eqs. \eqref{eq:nu_row} and \eqref{eq:mu_column} is $1$. The same applies to their corresponding strings of factor matrix indexes $\nu_i$ and $\mu_i$. We emphasize that one can set arbitrarily this lowest value to the integer $z$, whence $\eta$ and $\xi$ become $\eta_z$ and $\xi_z$, respectively, and so
					\begin{subequations}
					\label{eq:gen_eta_xi}
					\begin{align}
					\eta_z, \xi_z & : \underbrace{\mathbb{Z} \times \mathbb{Z} \times \cdots \times \mathbb{Z}}_{N} \longrightarrow \mathbb{Z}\\
					\nu & = \eta_z (\nu_1 , \ldots , \nu_N) = z + \sum^N_{i=1} \mathpzc{W}_{R,i} (\nu_i - z) \\
					\mu & = \xi_z (\mu_1 , \ldots , \mu_N) =  z + \sum^N_{i=1} \mathpzc{W}_{C,i} (\mu_i - z) \ .
					\end{align}
					\end{subequations}
In this article, we shall adopt the following compression maps: $\eta_0$ and $\xi_0$, \emph{i.e.} we set the lowest value of the indices $\nu, \nu_i,\mu,\mu_i$ to zero.

\par We may apply these results to the construction of finite tensor spaces. Say we have a collection of finite vector spaces $\{\mathcal{V}_1 , \ldots , \mathcal{V}_N\}$ from which we create the vector space $\mathcal{V}$ through the ordered tensor product
					\begin{equation}
					\label{eq:tensor_space_V}
					\mathcal{V} \equiv  \mathcal{V}_1 \otimes \cdots \otimes \mathcal{V}_{N-1}  \otimes  \mathcal{V}_N\ .
					\end{equation}
The basis of $\mathcal{V}$ is given by the ordered Kronecker product of the basis of the spaces $\{\mathcal{V}_1 , \ldots , \mathcal{V}_N\}$. Assume we number progressively (starting from $1$) the elements of the basis of each space $\mathcal{V}_i$, and then represent each basis element by a column vector. The basis $B_i$ of $\mathcal{V}_i$ may thus be represented, symbolically, by a column vector,
					\begin{equation}
					\label{eq:B_i}
					\mathpzc{B}_i = \begin{pmatrix}
					\gv{e}^{(i)}_1 \\
					\vdots \\
					\gv{e}^{(i)}_{d_i}
					\end{pmatrix}
					\end{equation}
where $\gv{e}^{(i)}_k$ is the assigned $k-$th basis element of the vector space $\mathcal{V}_i$, and $d_i = \dim \mathcal{V}_i$. Then, the following corollary follows from the index compression map theorem:
\begin{corollary}\label{corollary:row_index}
Let $\mathpzc{B}$ be the column representation of the basis of $\mathcal{V}$, Eq. \eqref{eq:tensor_space_V}, then 
					\begin{subequations}
					\label{eq:K_product_basis}
					\begin{align}
					\mathpzc{B} & =  \mathpzc{B}_1 \otimes \cdots \otimes \mathpzc{B}_{N-1} \otimes \mathpzc{B}_N  \\
					\gv{e}_\nu & = \gv{e}^{(1)}_{\nu_1} \otimes \cdots \otimes \gv{e}^{(N-1)}_{\nu_{N-1}}  \otimes \gv{e}^{(N)}_{\nu_N} \label{eq:nu_basis_vector}
					\end{align}
					\end{subequations}
where $\gv{e}_\nu$ is the $\nu-$th element of the basis of $\mathcal{V}$, and
					\begin{equation}
					\label{eq:nu_reducible}
					 \nu= \eta (\nu_1 , \ldots , \nu_N) = 1 + \sum^N_{i=1} \mathpzc{W}_{R,i} (\nu_i - 1) \ .
					\end{equation}
\end{corollary} 
Note that in writing Eqs. \eqref{eq:B_i} and \eqref{eq:K_product_basis} we have ignored the column indexes since they are always equal to 1. At this point, it is crucial we draw the reader's attention to the following: from Eq. \eqref{eq:K_product_basis} we see that to every tensor product of basis, we can associate a unique ordered string of the type $\{\nu_1 , \ldots  , \nu_{N-1},\nu_N \}$, which in turn corresponds to the unique basis $\gv{e}_\nu$, where $\nu$ is given by \eqref{eq:nu_reducible}. There is thus a one-to-one correspondence between the string $\{\nu_1 , \ldots  , \nu_{N-1},\nu_N \}$ and $\nu$, as already stated by the index compression map theorem (theorem \ref{thm:basis-numbers-mapping}). What this means is that we can forgo working directly with the tensor product of basis and work instead with $\nu \in \{1 , \ldots , \mathpzc{W}_{R,0}\}$ and its corresponding string $\{\nu_1 , \ldots  , \nu_{N-1},\nu_N \}$. The index $\nu$ in Eq. \eqref{eq:nu_reducible} is said to be a "reducible" index, while the indexes $\{\nu_1 , \ldots  , \nu_{N-1},\nu_N\}$ are its "irreducible" indexes (or components). It is worth noting that $\mathpzc{W}_{R,0}$, Eq. \eqref{eq:def_W_R},  coincides with the dimension of the tensor space $\mathcal{V}$. Note however that if, instead of the $\eta$, we employ the map $\eta_0$ in Eq. \eqref{eq:nu_reducible}, then $\nu \in \{0,1,\ldots,\mathpzc{W}_{R,0}-1\}$ and every irreducible index shall have $0$ as minimum instead of $1$.
\par The application of the index compression mappings in finite tensor Hilbert spaces is straightforward. We recall here again that in the following we shall be using the map $\eta_0$. The reason why will be explained shortly.

\section{The Holstein-Primakoff transformation}\label{sec:HP_transformation}
\par As it is well-known, in the uncoupled representation, the basis of the $i-$th spin Hilbert space $\mathcal{H}_i$ is $\{\ket{j_i, m_i}\}$, where $j_i$ is the spin quantum number and $m_i$ an eigenvalue of the operator  $\op{J}_{i,z}$. Certainly, the dimension of $\mathcal{H}_i$ is given by the number of possible values $m_i$ can take, which is $2j_i+1$. Since the basis $\{\ket{j_i, m_i}\}$ is a countable set, we may choose to assign an integer to each element of the set. There are numerous possible ways to do so but let us settle on the following: we start with the basis element with the highest value of $m_i$, \emph{i.e.} $\ket{j_i,j_i}$ and assign to it the integer $0$. In other words, we make the transformation $\ket{j_i,j_i} \mapsto \ket{0}$. We then take the next basis element $\ket{j_i,j_i-1}$ and assign to it the integer $1$. So $\ket{j_i,j_i-1} \mapsto \ket{1}$, and so forth -- up to $\ket{j_i,-j_i } \mapsto \ket{2j_i}$. In general, we have effected the transformation $\ket{j_i,m_i} \mapsto \ket{j_i-m_i}$. The new basis $\{\ket{0}, \ldots , \ket{2j_i} \}$ is the integer representation of $\mathcal{H}_i$'s basis and the integers define the irreducible indexes on $\mathcal{H}_i$. What we have just accomplished is nothing less than a second quantization. Precisely, it is part of what is termed in the literature as the "Holstein-Primakoff transformation".
\par  In concrete terms, the Holstein-Primakoff (HP) transformation\citep{book:Auerbach-1998,art:Holstein-1940} is a one-flavored mapping of spin-$j$ operators to boson creation and annihilation operators.  By "one-flavored" we mean only one type of boson is involved. Consider a collection $\mathpzc{A}$ of $N$ spins: $\mathpzc{A}=\{j_1 , j_2 \ldots , j_N\}$. Then, the HP transformation for the $i-$th spin operators reads
			\begin{subequations}
			\label{eq:HP_spin_transformations}
			\begin{align}
\hat{\mathscr{J}}^{-}_i & \mapsto \hat{b}^{\dagger}_i \ \left(2j_i \hat{\mathbf{1}}_i -\hat{b}^{\dagger}_i\hat{b}_i \right)^{1/2}  \\
\hat{\mathscr{J}}^{+}_i & \mapsto \left(2j_i \hat{\mathbf{1}}_i -\hat{b}^{\dagger}_i \hat{b}_i \right)^{1/2} \ \hat{b}_i \\
\hat{\mathscr{J}}^{z}_i & \mapsto j_i \hat{\mathbf{1}}_i - \hat{b}^{\dagger}_i\hat{b}_i
			\end{align}
			\end{subequations}
where as usual,
			\begin{equation}
			\hat{b}_i \ket{\nu}_i = \sqrt{\nu_i} \ket{\nu_i-1} \ , \ \hat{b}^\dagger_i \ket{\nu_i} = \sqrt{\nu_i+1} \ket{\nu_i+1} \ , \ \hat{b}^\dagger_i \hat{b}_i\ket{\nu_i}=\nu_i\ket{\nu_i}  \ , \ \left[\hat{b}_i,\hat{b}^\dagger_{i'} \right] = \delta_{i,i'}\hat{\mathbf{1}}_i
			\end{equation}
and $\nu_i \in \{0,1,2, \ldots , 2j_i \}$. The bosons in discussion here are the so-called Holstein-Primakoff bosons (or magnons), and one may think of the spins as vertexes of a complete graph -- where each vertex can accommodate not more than a certain fixed number of these particles\citep{art:Gyamfi-2018,art:Mendonca-2013}. For this reason, we shall sometimes refer to the spins as "spin vertexes" when working in the HP spin representation. Moreover, noteworthy is the observation that the occupation number of the HP bosons uniquely specifies the orientation of the spin along the axis of quantization. That is, independent of whether one is dealing with half-integral spins or not, any spin orientation is indicated by a nonnegative integer in the HP transformation. This is no trivial computational advantage. 
\par The transformations listed in Eq. \eqref{eq:HP_spin_transformations} ensure that the usual spin commutation relations are obeyed, namely
			\begin{equation}
			\left[ \hat{\mathscr{J}}^\alpha_i , \hat{\mathscr{J}}^\beta_{i'} \right]= i \epsilon^{\alpha \beta \gamma} \delta_{i,i'} \hat{\mathscr{J}}^{\gamma}_i  \ ,
			\end{equation}
where $\epsilon^{\alpha \beta \gamma}$ is the three dimensional Levi-Civita symbol.
\par There is thus a one-to-one correspondence between the basis $\{\ket{j_i,m_i}\}$ and the integer basis $\{\ket{\nu_i}\}$ under the HP transformation\citep{art:Gyamfi-2018}. In explicit terms,
			\begin{equation}
			\ket{j_i,m_i} \mapsto \frac{1}{\sqrt{(j_i-m_i)}!}\left( \hat{b}^\dagger_i \right)^{j_i-m_i} \ket{0} = \ket{\nu_i}
			\end{equation}
where $\nu_i=j_i-m_i$.
\par For what concerns magnetic resonance, the $N$ spins may consist of $N_{el}$ unpaired electrons and $N_{nu}$ nuclei of a given radical, for example. Then the (spin) Hilbert space $\mathcal{H}$ of the system may be defined according to the ordered Kronecker product
					\begin{equation}
					\label{eq:def_H_MULTI-SPIN}
					\mathcal{H} = \mathcal{H}_1 \otimes \cdots \otimes \mathcal{H}_{N-1} \otimes \mathcal{H}_N \ ,
					\end{equation}
where $\mathcal{H}_i$ represents the $i-$th spin's Hilbert space. Certainly, a viable basis $\mathpzc{B}$ for $\mathcal{H}$ consists of the Kronecker product (maintaining the same order as in Eq. \eqref{eq:def_H_MULTI-SPIN}) of the basis of the isolated $\mathcal{H}_i$:
			\begin{equation}
			\label{eq:basis_normal}
			\mathpzc{B} = \{ \ket{j_1,m_1} \otimes \ket{j_2,m_2} \otimes \cdots \otimes \ket{j_N, m_N} \} \ ,
			\end{equation}
where $-j_i \leq m_i \leq j_i $. This is obviously the uncoupled representation. Naturally, if we move to the integer representation, we may associate in a bijective manner a unique integer (reducible index, $\nu$) to each basis element of $\mathpzc{B}$, Eq. \eqref{eq:basis_normal}, employing the row index compression map $\eta_0$ (Corollary 1.1) . Thus, we have
					\begin{equation}
					\ket{\nu} = \ket{\nu_1}	\otimes \cdots \otimes \ket{\nu_{N-1}} \otimes \ket{\nu_N}	\equiv \ket{\nu_1, \ldots, \nu_{N-1},\  \nu_N}	\ ,
					\end{equation}	
where $\ket{\nu_i}$ is the integer representation of that basis element of $\mathcal{H}_i$ which contributes to the tensor product defining $\ket{\nu}$. Having chosen the mapping $\eta_0$, it follows from Eq. \eqref{eq:gen_eta_xi} that
					\begin{equation}
					\nu = \eta_0(\nu_1, \ldots , \nu_N) = \sum^N_{i=1} \mathpzc{W}_{R,i} \ \nu_i  \ ,
					\end{equation}
and $\nu_i \in \{0,1, \ldots, 2j_i\}$. It should now be clear why the choice $\eta_0$: with it, the enumeration of the normal spin basis elements coincides exactly with the possible occupation numbers of the HP bosons.
\par It must be emphasized that while the irreducible indexes $\{\nu_i\}$ indicate occupation numbers of the HP bosons, the reducible ones $\{ \nu \}$ enumerate the various possible permutations of the occupation numbers of these bosons.
\section{The row index compression map, the HP bosons and the diagonalization of static isotropic multispin Hamiltonians}\label{sec:Row_HP_diag}
\par We show in this section, as a matter of illustration, how the mapping $\eta_0$ together with the HP bosons can lead to the block diagonalization of an isotropic static multispin  Hamiltonian (\emph{i.e.} one with only isotropic coupling constants, $g-$factors included), and thus to a diagonalization scheme which is computationally more affordable. In the following, all bases mentioned are to be understood as orthonormal.
\par We now subject the spin multiset $\mathpzc{A}$ to a magnetic field of intensity $B_0$ applied along the $z-$axis, which is chosen as the quantization axis. The isotropic spin Hamiltonian $\op{H}_{spin}$ of the system in the uncoupled representation is then
					\begin{equation}
					\label{eq:H_spin scalar-like}
					\op{H}_{spin} =  B_0 \sum_{i}  \mu_i \ g_{i} \op{J}^z_{i} + \sum_{i > i'} T_{i, i'} \  \op{J}_{i} \cdot \op{J}_{i'} 
					\end{equation}
where $\op{J}_i\  \left(=  \op{J}^x_i + \op{J}^y_i + \op{J}^z_i\right)$, $\mu_i$ and $g_i$ are the spin operator, the appropriate magneton (Bohr or nucleus) and the (isotropic) $g-$factor of the $i-$th spin, respectively, and $T_{i, i'}$ is the coupling constant  (for example, the dipole-dipole coupling constant plus others) of the interaction between spin $i$ and $i'$. $\op{H}_{spin}$ may be written explicitly in matrix representation as an operator $\mathfrak{G}_{spin}$ acting on the identity operator $\mathbb{I}^{\otimes N}$ of $\mathcal{H}$, \emph{i.e.}\citep{inbook:Barone_et_al-2016}
					\begin{equation}
					\label{eq:spin Hamiltonian}
					\mathscr{H}_{spin} 
					= \mathfrak{G}_{spin} \ \mathbb{I}^{\otimes N}
					\end{equation}				
with,
\begin{subequations}
\begin{align}
\mathbb{I}^{\otimes N} & := \bigotimes^N_{i=1} \mathbb{I}_{i} = \mathbb{I}_{i=1} \otimes \cdots \otimes \mathbb{I}_{i=N} \\
 \mathfrak{G}_{spin} & :=  B_0 \sum_i \mathfrak{G}\left(\mathbb{I}_{i},  \mu_i g_{i} \op{J}^z_{i}\right)  + \sum_{i,i'( > i)} \sum_{\beta \in \{x,y,z\}} T_{i, i'} \mathfrak{G}\left(\mathbb{I}_{i}, \op{J}^\beta_{i}; \mathbb{I}_{i'}, \op{J}^{\beta'}_{i'}\right) \label{eqn:multi-spin_Ham_3} \ ,
 \end{align}
\end{subequations}
where $\op{J}^{\beta}_{i}$ is the spin angular momentum operator of the $i-$th spin along the axis $\beta$, and where $\mathfrak{G}$ is the substitution operator\citep{inbook:Barone_et_al-2016} defined operationally as
					\begin{equation}
					 \mathfrak{G}(x'_1,x_1;\ldots ;x'_n,x_n) f(x_1, \ldots , x_n) = f(x'_1, \ldots , x'_n) \ .
					\end{equation}
The steady increase in $\op{H}_{spin}$'s sparseness is evident in Eq. \eqref{eq:spin Hamiltonian}, if we notice that $\mathfrak{G}_{spin}$ comprises substitution operators which substitute at most two identity matrices of $\mathbb{I}^{\otimes N}$ with other matrices, independent of the total number of spins $N$.

\subsection{Block diagonalizing the spin Hamiltonian}\label{sec:block-diag_spin_Hamil}
\par For starters, we rewrite $\op{H}_{spin}$ according to the HP transformation:
				\begin{multline}
				\label{eq:spin_Hamiltonian_HP}
				\op{H}_{spin} =  B_0 \sum_i \mu_i g_i \left(j_i - \hat{b}^\dagger_i \hat{b}_i \right) + \sum_{i > i'} T_{i,i'}\left(j_i - \hat{b}^\dagger_i \hat{b}_i \right)\left(j_{i'} - \hat{b}^\dagger_{i'} \hat{b}_{i'} \right) \\
				+ \frac{1}{2} \sum_{i > i'} T_{i,i'} \left[\hat{b}^\dagger_i \left(2j_i - \hat{b}^\dagger_i \hat{b}_i\right)^{1/2} \left(2j_{i'} - \hat{b}^\dagger_{i'} \hat{b}_{i'}\right)^{1/2} \hat{b}_{i'} + h.c.\right] \ .
				\end{multline}
\par Consider now the matrix element $\matrixel{\lambda}{\op{H}_{spin}}{\nu}$ between two arbitrary reducible indexes, $\lambda$ and $\nu$, of $\mathcal{H}$. It follows from Eq. \eqref{eq:spin_Hamiltonian_HP} that
%					\widetext
					\begin{multline}
					\label{eq:spin_hamil}
					\matrixel{\lambda}{\op{H}_{spin}}{\nu} =  \delta_{\lambda \hspace{0.04cm}, \nu} \left[ B_0 \mu_B \sum^N_{i = 1}    g_i  (j_i - \nu_i) +  \sum_{i > i'} T_{i , i'} \ (j_i - \nu_i) (j_{i'} - \nu_{i'})\right] \\ 
					+ \frac{1}{4}\sum_{i>k} T_{i , k} \left\lbrace \left(\prod_{l \neq i , k} \delta_{\lambda_l, \nu_l} \right)  \  \delta_{\lambda_i, \nu_{i}  \pm 1} \ \delta_{\lambda_k , \nu_{k} \pm 1} \Big[ 1 - \text{sgn}\left(\lambda_i - \nu_i \right) \text{sgn}\left(\lambda_k - \nu_k \right) \Big] \right. \\
					\times   s_{\pm}\left(\nu_i \right)s_{\pm}\left(\nu_k \right) \Bigg\}  
					\end{multline}						
where 
					\begin{equation}
					s_{\pm}(\nu_i):= \sqrt{\left(\nu_i + \frac{1 \mp 1}{2}  \right)\left(2j_i- \nu_i + \frac{1 \pm 1}{2}  \right)} \ .
					\end{equation}
Notice that the matrix element $\matrixel{\lambda}{\mathscr{H}_{spin}}{\nu}$ reduces to only one of the two terms in Eq. \eqref{eq:spin_hamil}: it reduces to the first term if the two reducible indexes $\lambda$ and $\nu$ coincide, otherwise it yields the second term. 
\par Regarding the second term, one observes that the necessary and sufficient condition (assuming $T_{i , k} \neq 0$) for any of its summand to be non nonzero is if the following conditions are both met:
	\begin{itemize}
	\item if of the $N$ spin vertexes, there is no variation in the number of bosons occupying $N-2$ spin vertexes (\emph{i.e.} we have $N-2$ invariant irreducible indexes) as the transition from the initial reducible index $\nu$ to the final index $\lambda$ is made;
	\item if of the remaining two spin vertexes, the transition from $\nu$ to $\lambda$ is such that one loses a boson to the other.
\end{itemize}	 
These observations actually suggest a good quantum number in the chosen representation; namely, \emph{the sum of the irreducible components of the reducible indexes constitute good quantum numbers}. In other words, \emph{the total number of HP bosons are conserved} during transitions. Indeed, the total boson occupation number operator $\op{N}=\sum^N_{i=1} \hat{b}^\dagger_i \hat{b}_i$ commutes with $\op{H}_{spin}$. Therefore, a necessary but not sufficient condition for $\matrixel{\lambda}{\op{H}_{spin}}{\nu}$ to be nonzero is that
						\begin{subequations}
						\label{eq:total_boson_constraint}
						\begin{align}
						n(\nu) & = n(\lambda) \\
						n(\nu) & := \matrixel{\nu}{\op{N}}{\nu}=\sum^N_{i=1} \ \nu_i  \ .	\label{eq:n_composition}				
						\end{align}
						\end{subequations}
\par Say $\mathpzc{N}$ the set $\{n(\nu)\}$, \emph{i.e.} the set of all possible eigenvalues of the operator $\op{N}$. Since $n$ is a good quantum number of $\op{N}$, it follows that all reducible indexes with the same $n$ form a subspace of the total spin Hilbert space $\mathcal{H}$. And any two subspaces characterized by two different $n$ values are orthogonal to each other. It is evident that
				\begin{equation}
				\mathpzc{N} \equiv \left\lbrace 0, 1, \dots , 2J_0 \right\rbrace \ ,
				\end{equation}
where $J_0 = \sum_i j_i$. Consequence of these considerations is the following proposition:
\begin{prop}
Given a collection of $N$ spins with spin quantum numbers $j_1, \ldots , j_N$ whose interactions are described by an isotropic static spin Hamiltonian of the form in Eq. \eqref{eq:spin_hamil}, the block-diagonalized $\mathscr{H}_{spin}$ consists of exactly $2J_0+1$ submatrices, \emph{i.e.}
				\begin{equation}
				\label{eq:block-diag_H_SPIN}
				\mathscr{H}_{spin} = \bigoplus^{2J_0}_{n=0} \mathscr{B}_n = \diag \left(\mathscr{B}_0, \mathscr{B}_1, \ldots , \mathscr{B}_{2J_0} \right)  \ . 
				\end{equation}						
\end{prop}
\par The complexity of finding the eigenvalues and eigenvectors of $\mathscr{H}_{spin}$ thus reduces to that of finding the same for $2J_0+1$ smaller matrices. In particular, if $\mathscr{U}_{spin}$ and $\mathscr{D}_{spin}$ are $\mathscr{H}_{spin}$'s eigenvector and eigenvalue matrices, respectively, then 
				\begin{equation}
				\label{eq:U_spin_matrix}
				\mathscr{U}_{spin} = \bigoplus^{2J_0}_{n=0} \mathscr{U}_n \qquad \mbox{and} \qquad \mathscr{D}_{spin} = \bigoplus^{2J_0}_{n=0} \mathscr{D}_n
				\end{equation}				 
where $\mathscr{U}_n$ and $\mathscr{D}_n$ are the matrices composed of the eigenvectors and eigenvalues of the $n-$th submatrix in Eq. \eqref{eq:block-diag_H_SPIN}, respectively.
\par It is worth noting from Eq. \eqref{eq:block-diag_H_SPIN} that the total number of submatrices $\mathscr{B}_n$ is determined by the total spin $J_0$ and not by the total number of spins $N$. This translates into the implication that any arbitrary collection of spins having the same total spin number $J_0$ described by the Hamiltonian $\op{H}_{spin}$ of Eq. \eqref{eq:spin_hamil} will have the same number of submatrices when the latter is block-diagonalized, though their dimension may differ from system to system.

\subsection{Dimension of the submatrices. Density and sparseness of $\mathscr{H}_{spin}$} \label{sec:dimension_block_matrices}
\par The use of HP bosons allow an easy analytical characterization of the total spin Hilbert space $\mathcal{H}$. Since all reducible indexes with the same number of HP bosons $n$ span the same subspace of $\mathcal{H}$, the dimension of the submatrix $\mathscr{B}_n$, $\Omega_{\mathpzc{A},n}$, is simply the number of compositions of the integer $n$ according to Eq. \eqref{eq:n_composition}. We recall that the irreducible indexes $\nu_i$ in the mentioned equation are subject to the constraint $0 \leq \nu_i \leq 2j_i$. 
\par The analytical determination of $\Omega_{\mathpzc{A},n}$ is an enumerative combinatoric problem\cite{art:Gyamfi-2018}. In particular, it can be proved that the generating function, $G_{\mathpzc{A},\Omega}(q)$, for the integers $\{\Omega_{\mathpzc{A},n}\}$ is of the form\cite{art:Gyamfi-2018}:
				\begin{equation}
				\label{eq:generating_function_G}
				G_{\mathpzc{A},\Omega}(q) = \prod^\sigma_{\alpha=1} \left( \left[2j_\alpha + 1 \right]_q \right)^{N_{\alpha}} = \sum^{2J_0}_{n=0} \ \Omega_{\mathpzc{A},n} \ q^n \ ,
				\end{equation}
where $\left[ m \right]_q$ simply indicates the $q-$analogue of the integer $m$, \emph{viz.}
				\begin{equation}
				\left[ m \right]_q := \frac{1-q^{m}}{1-q} = 1 + q + q^2 + \ldots + q^{m-1} \ .
				\end{equation}
The dimension of the submatrices may be computed straightforward analytically by resorting to the following formula\cite{art:Gyamfi-2018}
				\begin{multline}
				\label{eq:generalized_dim_B(n)}
				\Omega_{\mathpzc{A},n} = \sum_{\substack{\sum^{\sigma}_{\alpha = 1} (2j_\alpha + 1)s_\alpha \leq n \\ 0 \leq s_\alpha \leq N_\alpha}} (-1)^{s_1 + \ldots + s_\sigma} 
				\binom{
				N + n - 1 - \sum^{\sigma}_{\alpha = 1} (2j_\alpha + 1)s_\alpha}{
				N- 1
				} 
				\binom{
				N_1}{
				s_1
				}\ldots
				\binom{
				N_\sigma }{
				s_\sigma} \ .
				\end{multline}
We remark that the coefficients $\{\Omega_{\mathpzc{A},n}\}$ are reciprocal and unimodal, namely, %
				\begin{subequations}
				\begin{align}
				\Omega_{\mathpzc{A}, n} & = \Omega_{\mathpzc{A}, 2J_0-n} \ , \ \qquad \forall n \in \mathpzc{N} \\
				\Omega_{\mathpzc{A},n} & \leq \Omega_{\mathpzc{A},n'}, \ \mbox{for  $n< n'$ and $ 0\leq \{n,n'\} \leq \lfloor{J_0\rfloor}$} \ ,
				\end{align}
				\end{subequations}
respectively. In addition, $\max \{\Omega_{\mathpzc{A}, n}\}= \Omega_{\mathpzc{A}, \lfloor J_0 \rfloor}$, where $\lfloor \bullet \rfloor$ is the floor function.
\par Interestingly, we note from Eq. \eqref{eq:generalized_dim_B(n)} that $\Omega_{\mathpzc{A},n}$ -- being the sum of products of binomial coefficients -- implies that the computational cost of diagonalizing the $\{\mathscr{B}_n\}$ submatrices is still exponential in $N$ but scales lesser with the latter as compared to diagonalizing the whole Hamiltonian at once. As an example, consider $10$ spin$-1/2$s described by a static $\op{H}_{spin}$ of the form in Eq. \eqref{eq:spin_hamil}. The dimension of its global Hilbert space is $2^{10} = 1024$, which means finding its eigenvalues and eigenvectors would require diagonalizing a matrix of dimension $1024$. But using the row index compression map $\eta_0$ and the HP bosons, the same eigenvectors and eigenvalues can be determined by diagonalizing eleven (in reality only nine) matrices, the largest being of dimension $252$. In particular, submatrices of dimension $1,10,45,120$ and $210$ will appear in pairs, and only one will be of dimension $252$. If we now add another spin-$1/2$ to the system, we will have to diagonalize now a matrix of dimension $2048$ but employing the scheme explained above we will need to diagonalize smaller matrices of dimension not greater than $462$. 
\par Still on the topic of analytic characterization of the multispin Hamiltonian, we may study for example the density of $\mathscr{H}_{spin}$. If we define the density $\zeta(A)$ of the matrix $A$ as the ratio between its number of nonzero elements and the total number of entries, then it can be proved that given a collection of $N$ spins and assuming a multispin Hamiltonian of the form \eqref{eq:H_spin scalar-like},  the density of $\mathscr{H}_{spin}$ has an upper bound given by the expression
					\begin{equation}
					\label{eq:upper_bound_zeta_HS}
					\sup_{\zeta \in \mathcal{Z}\left[\wp\left(\mathscr{H}_{spin}\right)\right]} \zeta \left(\mathscr{H}_{spin}\right) =\frac{\sum^{2J_0}_{n=0} \Omega^2_{\mathpzc{A},n}}{\prod^\sigma_{\alpha=1} (2j_\alpha +1)^{2N_\alpha}} \ ,
					\end{equation}
where $\wp\left(\mathscr{H}_{spin}\right)$ is the order\cite{inbook:Barone_et_al-2016} of $\mathscr{H}_{spin}$ and $\mathcal{Z}\left[\wp\left(\mathscr{H}_{spin}\right)\right]$ is the set of all possible $\zeta$s of $\mathscr{H}_{spin}$. In the limit case whereby the spin multiset $\mathpzc{A}$ consists of only spin-1/2s, \emph{i.e.} $\mathpzc{A}=\left\lbrace \frac{1}{2}^N \right\rbrace$, one derives from Eq. \eqref{eq:upper_bound_zeta_HS} that 
					\begin{equation}
					\zeta \left(\mathscr{H}_{spin}\right) \leq 2^{-2N}\binom{2N}{N} \ ,
					\end{equation}
which for very large $N$ reads approximately
					\begin{equation}
					\zeta \left(\mathscr{H}_{spin}\right) \lesssim \frac{1}{\sqrt{\pi N}} \ . 
					\end{equation}				
The density of the spin Liouvillian $\mathfrak{L}_{spin}$ on the other hand has an upper bound of
					\begin{equation}
					\label{eq:upper_bound_zeta_LS}
					\sup_{\zeta \in \mathcal{Z}\left[\wp\left(\mathfrak{L}_{spin}\right)\right]} \zeta \left(\mathfrak{L}_{spin}\right) =\frac{\left(\sum^{2J_0}_{n=0} \Omega^2_{\mathpzc{A},n} \right)^2}{\prod^\sigma_{\alpha=1} (2j_\alpha +1)^{4N_\alpha}} \ .
					\end{equation}
which is the square of that of $\mathscr{H}_{spin}$. 
\par The parameter $\zeta$ is a good measure of the fraction of the total Hilbert (or Liouville) space which effectively encodes the whole spin dynamics. A closely related measure is the sparseness $\chi := 1- \zeta$. 

\subsection{Exploiting $T$-symmetry}
Although the spin Hamiltonian, Eq. \eqref{eq:H_spin scalar-like}, we are concerned with here is real, it fails to be time-reversal (or $T-$symmetry) invariant (in the conventional sense) due to the presence of the Zeeman term. Indeed, as it is well-known, the presence of an external magnetic field in a non-spinless system always leads to broken $T-$symmetry\cite{book:Haake-2001}. Nonetheless, the subspaces $\mathscr{B}_n$ and $\mathscr{B}_{2J_0-n}$ are related through $T-$symmetry. We show below how this relation could be exploited to further reduce the computational cost of the numerical simulation of the magnetic resonance spectra of multispin systems.
\par Let $\op{T}_i$ represent the time-reversal operator acting on the $i-$th spin. Then, in the integer representation, the irreducible index $\ket{\nu_i}$ is transformed by $\op{T}_i$ as follows
			\begin{equation}
			\op{T}_i \ket{\nu_i} = \ket{2j_i - \nu_i} \equiv \ket{\overline{\nu}_i} \ , 
			\end{equation}
which means the unitary part of $\op{T}_i$ (defined up to a phase factor) rotates the $\{\ket{\nu_i}\}$ vectors by an angle of $\pi$ around the $x-$axis. The vector $\ket{\overline{\nu}_i}$ is said to be the \emph{hole complement} of $\ket{\nu_i}$. As a matter of fact, in the HP representation, the time-reversal symmetry could be interpreted as the particle-hole symmetry.

 Since 
			\begin{equation}
			\op{T}^2_i \ket{\nu_i} = \ket{\nu_i}
			\end{equation}
we have that $\op{T}^2_i=\hat{\mathbf{1}}_i \Longrightarrow \op{T}_i= \op{T}_i^{\dagger}$. \par For a multispin system, the total time-reversal operator reads
			\begin{equation}
			\op{T} = \prod_i \op{T}_i
			\end{equation}
and
			\begin{equation}
			\label{eq:particle-hole_symm}
			\begin{split}
			\op{T}\ket{\nu}& =\op{T} \ket{\nu_1 , \nu_2 , \ldots, \nu_N} \\
			& = \ket{2j_1-\nu_1, 2j_2-\nu_2, \ldots , 2j_N-\nu_N} \\
			& = \ket{\mathpzc{W}_{R,0}-1-\nu} := \ket{\overline{\nu}}\ .
			\end{split}
			\end{equation}
$\ket{\overline{\nu}}$ is clearly the hole complement of $\ket{\nu}$. Naturally, $\op{T}=\op{T}^{\dagger}$. 
\par We now investigate how $\op{T}$ rotates $\op{H}_{spin}$. 
To begin with, it can be easily proved that
			\begin{subequations}
			\label{eq:T-symm_trans_no_op_plus_ladders}
			\begin{align}
			\op{T}_i\hat{b}^\dagger_i \hat{b}_i \op{T}_i^{\dagger} & = 2j_i-\hat{b}^\dagger_i \hat{b}_i \ ,\\
			\op{T}_i\op{J}^{\pm}_i \op{T}^{\dagger}_i & = \op{J}^\mp_i \ .
			\end{align}
			\end{subequations}
Then, as a consequence of the transformations in Eqs. \eqref{eq:T-symm_trans_no_op_plus_ladders}, we are led to the following important identity,
			\begin{equation}
			\label{eq:T-symm_H_spin_transformation}
			\op{H}_{spin} - \op{T} \op{H}_{spin} \op{T}^{\dagger} = 2 B_0 \sum_i \mu_ig_i \left(j_i - \hat{b}^\dagger_i \hat{b}_i \right) \ .
			\end{equation}
That is, the spin Hamiltonian $\op{H}_{spin}$ and its time-reversally rotated counterpart differ by twice the Zeeman term, which is perfectly in line with what we should expect since the spin-spin interaction terms in $\op{H}_{spin}$ are $T-$invariant.
\par To understand the importance of Eq. \eqref{eq:T-symm_H_spin_transformation}, we must bear in mind that if $\ket{\lambda}$ and $\ket{\nu}$ are elements of the basis which span the subspace $\mathscr{B}_n$, then their complements $\ket{\overline{\lambda}}$ and $\ket{\overline{\nu}}$ also span $\mathscr{B}_{2J_0-n}$. It follows from Eq. \eqref{eq:T-symm_H_spin_transformation} that
			\begin{equation}
			\label{eq:T-symmetry_matrix_repre_0}
			\matrixel{\lambda}{\op{H}_{spin}}{\nu} - \matrixel{\overline{\lambda}}{\op{H}_{spin}}{\overline{\nu}} = \delta_{\lambda , \nu}\  2 B_0 \sum_i \mu_i g_i \left(j_i - \nu_i \right) \ ,
			\end{equation}
which greatly simplifies the determination of the matrix elements of $\mathscr{B}_{2J_0-n}$ if those of $\mathscr{B}_n$ are known. In concrete terms, $\mathscr{B}_{2J_0-n}$ and $\mathscr{B}_n$ differ only in their diagonal elements, the difference given by twice the corresponding Zeeman terms of $\mathscr{B}_n$ . In terms of matrix representation, we have
			\begin{equation}
			\label{eq:T-symmetry_matrix_repre}
			\mathscr{B}_{2J_0-n} = \mathscr{T}\mathscr{B}_{n} \mathscr{T} - 2 Z_n  \qquad \forall \ n \in \mathpzc{N}
			\end{equation}
where $Z_n$ is the matrix representation of the Zeeman term in Eq. \eqref{eq:H_spin scalar-like} for the subspace $\mathscr{B}_n$, and $\mathscr{T}$ in  Eq. \eqref{eq:T-symmetry_matrix_repre} could be understood, without loss of generality, as the unit antidiagonal matrix of order $\Omega_{\mathpzc{A},n}$. A straight-away computation of the eigenvalues and eigenvectors of $\mathscr{B}_{2J_0-n}$ when those of $\mathscr{B}_{n}$ are known could be particularly interesting and useful.

\subsection{Eigenvalues and eigenvectors of the Liouvillian $\sop{L}_{spin}$}	
It mostly happens that in theoretical derivations, working in Liouville space appears to be more convenient. We may say that the Liouville space is also ideal for simulating magnetic resonance spectra, but workers in the field are well acquainted with the fact that from a practical point of view it is not since the dimension of the Liouville space is that of the corresponding Hilbert space squared, which means the computational cost gets roughly squared with respect to that in the Hilbert space if one tries to diagonalize the Liouvillian for subsequent calculations. This hurdle can be easily overcome as we explain below.
\par We recall that the matrix representation of the Liouvillian superoperator $\sop{L}_{spin}$ corresponding to $\op{H}_{spin}$ is
				\begin{equation}
				\mathfrak{L}_{spin} = \mathscr{H}_{spin} \otimes \mathbb{I}_{\mathcal{H}} - \mathbb{I}_{\mathcal{H}} \otimes \mathscr{H}_{spin} \ ,
				\end{equation}				 		
where $\mathbb{I}_{\mathcal{H}}$ is the matrix representation of $\mathcal{H}$'s identity operator. It can easily be proved that the supermatrix $\mathfrak{U}_{spin}$ which diagonalizes $\mathfrak{L}_{spin}$ is related to the eigenvector matrix of $\mathscr{H}_{spin}$, $\mathscr{U}_{spin}$, Eq. \eqref{eq:U_spin_matrix}, as
				\begin{equation}
				\label{eq:supermatrix_U}
				\mathfrak{U}_{spin} = \mathscr{U}_{spin} \otimes \mathscr{U}_{spin} \ ,
				\end{equation}
and the diagonal matrix of $\mathfrak{L}_{spin}$, $\mathfrak{D}_{spin}$, is related to $\mathscr{D}_{spin}$, Eq. \eqref{eq:U_spin_matrix}, through the identity
				\begin{equation}
				\label{eq:superdiagonal_D}
				\mathfrak{D}_{spin} = \mathscr{D}_{spin} \otimes \mathbb{I}_{\mathcal{H}} - \mathbb{I}_{\mathcal{H}} \otimes \mathscr{D}_{spin} \ .				
				\end{equation}
These last two relations are valid in general for real (Hermitian) Hamiltonians, independent of whether the Hamiltonian is time-dependent or not. For complex Hermitian Hamiltonians, they slightly change\footnote{In the general case, the Liouvillian $\mathfrak{L}$ of the complex Hamiltonian $\mathscr{H}$ defined on the finite Hilbert space $\mathcal{H}$, is 
				\begin{equation}
				\label{eq:general_Liouville}
				\mathfrak{L} = \mathscr{H} \otimes \mathbb{I}_{\mathcal{H}}   - \mathbb{I}_{\mathcal{H}} \otimes \mathscr{H}^T \ ,
				\end{equation}
where $\mathscr{H}^T$ is the transpose of $\mathscr{H}$. If $\mathscr{U} (\mathfrak{U})$ and $\mathscr{D} (\mathfrak{D})$ are the eigenvector and eigenvalue matrices, respectively, of $\mathscr{H} (\mathfrak{L})$, then
				\begin{equation}
				\begin{split}
				\mathfrak{U} & = \mathscr{U} \otimes \left(\mathscr{U}^{-1}\right)^T \ , \\
				\mathfrak{D} & = \mathscr{D} \otimes \mathbb{I}_{\mathcal{H}}   - \mathbb{I}_{\mathcal{H}} \otimes \mathscr{D} \ .
				\end{split}
				\end{equation}						
}.
\par Therefore, one understands that if the eigenvalues and eigenvectors of $\mathscr{H}_{spin}$ are known, those of the corresponding Liouvillian matrix $\mathfrak{L}_{spin}$ can be directly computed without having to diagonalize the superoperator de novo. The significance of the relations \eqref{eq:supermatrix_U} and \eqref{eq:superdiagonal_D} lies in the fact that the computational cost of finding the eigenvalues and eigenvectors of $\sop{L}_{spin}$ is now essentially the same as finding those of $\op{H}_{spin}$. So, with a diagonalization protocol of $\mathscr{H}_{spin}$ like the one presented in this article  whose computational cost scales very less exponentially with $N$, Eqs. \eqref{eq:supermatrix_U} and \eqref{eq:superdiagonal_D} allow simulation in Liouville space without much extra computational cost. This point is of great consequence because if we consider again the case of $10$ spin-$1/2$s, the Liouvillian $\mathfrak{L}_{spin}$ is a square matrix of dimension $1,048,576 $ but through Eqs. \eqref{eq:supermatrix_U} and \eqref{eq:superdiagonal_D}, its eigenvectors and eigenvalues can be determined by diagonalizing a series of matrices ($11$ in total) of dimension $1,10,45,120,210$ and $252$ using the scheme presented above. An additional relevance of the Eqs. \eqref{eq:supermatrix_U} and \eqref{eq:superdiagonal_D} has to do with the fact that they may allow an efficient perturbative computation of the resonance spectrum of the spin system in Liouville space inasmuch as they drastically reduce the computational cost of diagonalizing $\mathfrak{L}_{spin}$. 
\par We conclude this section pointing out that there is a one-to-one correspondence between the total occupation numbers $\{n\}$ and the set of all possible total spin projections along $z$ (or weights), \emph{i.e.} $\{M_n\}$\cite{art:Gyamfi-2018}. In particular, $M_n = J_0 - n$. Since the occupation numbers of the HP bosons relate to spin orientations, saying the number of HP bosons is conserved implies the total spin weights are also conserved:   thus $\Delta M=0$, for a unique $M_n$ --  hence we are dealing with zero-quantum transitions. The observation just made seems to disparage the importance of the diagonalization scheme presented above in relation to the study of multispin dynamics but that should not be the case for the following reason: $\op{H}_{spin}$ as given in Eq. \eqref{eq:spin_hamil} may be viewed as the multispin Hamiltonian in some convenient frame of reference, for instance, the molecular frame. Then, upon diagonalizing it, as we have shown above, one can perform a time-dependent (possibly stochastic) rotation of the equations describing the dynamics to that of the laboratory frame. And this will inevitably generate a dissipative term which will induce the transitions observed experimentally. The Hamiltonian in the non-dissipative term of the rotated equation of motion will have the same eigenvalues as $\op{H}_{spin}$, but the new eigenvectors will be simply rotated with respect to the original ones and can be easily determined. The latter may be used as the basis in some perturbative calculation of the contribution of the dissipative term to the evolution of the system under some appropriate assumptions. The rotation may also be directly performed in Liouville space effortlessly. The important thing noteworthy here is that the procedure we just described begins with the diagonalization of the static spin Hamiltonian in the chosen convenient frame. We shall return to this procedure in a future article.

\section{Application: the \ce{^{.}CH2OD} radical, a case study.}\label{sec:Application}	
Suppose we want to simulate the electron paramagnetic resonance (EPR) spectrum of the deuterated hydroxymethyl radical (\ce{^{.}CH2OD}). We will have to write its spin Hamiltonian in some basis. We assume the experimental conditions are such that the isotropic spin Hamiltonian given in Eq. \eqref{eq:H_spin scalar-like} is suitable, with all relevant constants known. Our aim here is not to work out the final spectrum but to show the usefulness of the mapping $\eta_0$ and the HP bosons. 
\par We have four species of nonzero spin in the \ce{^{.}CH2OD} radical: the unpaired electron (spin-$1/2$), the two methyl hydrogen atoms (each of spin-$1/2$) and the deuterium isotope (spin-$1$). The spin multiset $\mathpzc{A}$ we are considering here is therefore $\mathpzc{A}=\left\lbrace \frac{1}{2}^3, 1 \right\rbrace$. And the Hilbert space of the radical is thus
				\begin{equation}
				\label{eq:CH2OD_Hilbert_space}
				\mathcal{H} = \mathcal{H}^{(e)} \otimes \mathcal{H}^{(1)} \otimes \mathcal{H}^{(2)} \otimes \mathcal{H}^{(D)} \ ,
				\end{equation}
where $\mathcal{H}^{(e)}$ is the Hilbert space of the unpaired electron, while $\mathcal{H}^{(1)}, \mathcal{H}^{(2)}$ and $\mathcal{H}^{(D)}$ represent the Hilbert space of the two methyl hydrogens and deuterium, respectively. In writing \eqref{eq:CH2OD_Hilbert_space} we have chosen a specific order, which is of course not unique. Nonetheless, we must stick to the chosen order when constructing $\mathcal{H}$'s basis in the uncoupled representation. 
\par Now, in the integer representation, we know that all uncoupled states $\ket{j_i, m_i}$ undergo the transformation
				\begin{equation}
				\ket{j_i, m_i} \mapsto \ket{j_i-m_i}
				\end{equation}
where the nonnegative integer $(j_i-m_i)$ is the number of HP bosons occupying that spin vertex. For example, for the spin-$1/2$ particles we have
				\begin{equation}
				\ket{1/2, 1/2} \mapsto \ket{0} \ , \qquad \ \ket{1/2, -1/2} \mapsto \ket{1} \ ,
				\end{equation}
while for the spin-$1$ species we have
				\begin{equation}
				\ket{1, 1} \mapsto \ket{0} \ , \qquad \ \ket{1, 0} \mapsto \ket{1}  \ , \qquad \ \ket{1, -1} \mapsto \ket{2} \ .
				\end{equation}
The Hilbert space $\mathcal{H}$ of the radical is of dimension $24$, and each of its $24$ basis elements is specified by the tensor product involving one basis element of each of its $4$ subspaces. $\mathcal{H}$'s basis may be represented conveniently as (following the same order as in Eq. \eqref{eq:CH2OD_Hilbert_space}),
				\begin{equation}
				\label{eq:CH2OD_Hilbert_space_basis}
				\overbrace{
				\begin{pmatrix}
				\rket{0} \\
				\rket{1} \\
				\vdots \\
				\rket{23}
				\end{pmatrix}}^{\rket{\nu}} = 
				\overbrace{
				\begin{pmatrix}
				\ket{0} \\
				\ket{1}
				\end{pmatrix}}^{\ket{\nu_1}} \otimes 
				\overbrace{
				\begin{pmatrix}
				\ket{0} \\
				\ket{1}
				\end{pmatrix}}^{\ket{\nu_2}} \otimes 
				\overbrace{
				\begin{pmatrix}
				\ket{0} \\
				\ket{1}
				\end{pmatrix}}^{\ket{\nu_3}} \otimes 
				\overbrace{
				\begin{pmatrix}
				\ket{0} \\
				\ket{1} \\
				\ket{2}
				\end{pmatrix}}^{\ket{\nu_4}} \ . 				
				\end{equation}
where for the sake of clarity, we have indicated the basis elements of $\mathcal{H}$ in the integer representation as $\rket{\nu}$. It is clear from Eq. \eqref{eq:CH2OD_Hilbert_space_basis} that to each of the kets on the left hand side is associated a unique tensor product of the basis elements of the subspaces, thus the indices $\rket{\nu}$ of $\mathcal{H}$ are reducible. For example, $\rket{1} = \ket{0} \otimes \ket{0} \otimes \ket{0} \otimes \ket{1}= \ket{0,0,0,1}$. In matrix terms, that corresponds to the following column vector:
				\begin{equation}
				\rket{1} =  \begin{pmatrix}
				1 \\
				0
				\end{pmatrix} \otimes 
				\begin{pmatrix}
				1 \\
				0
				\end{pmatrix} \otimes
				\begin{pmatrix}
				1 \\
				0
				\end{pmatrix} \otimes
				\begin{pmatrix}
				0 \\
				1 \\
				0
				\end{pmatrix}
				\end{equation}
As mentioned earlier, the reducible indexes $\{\nu\}$ enumerate the various possible occupations of the HP bosons on the graph of spin vertexes. For instance, $\rket{1} =  \ket{0,0,0,1}$ relates to the HP boson configuration whereby all the spin-$1/2$ vertexes are empty and only the spin$-1$ vertex is occupied by a single HP boson.
\par The row-index compression map $\eta_0$ which relates the string $\{\nu_1, \nu_2, \nu_3, \nu_4\}$ to $\nu$ for the chosen order of the Kronecker product in Eq. \eqref{eq:CH2OD_Hilbert_space} is
				\begin{equation}
				\label{eq:CH2OD_nu}
				\nu = \eta_0(\nu_1, \nu_2,\nu_3,\nu_4) = 12\nu_1 + 6 \nu_2 + 3\nu_3 + \nu_4 \ .
				\end{equation}
From this equation, one can determine the reducible index $\nu$ for any given string $\nu_1, \nu_2, \nu_3, \nu_4$, and vice versa. For example, one can easily determine that $\rket{15}=\ket{1} \otimes \ket{0} \otimes \ket{1} \otimes \ket{0}$.
\par Having assumed that the Hamiltonian is of the form in Eq. \eqref{eq:H_spin scalar-like}, we are immediately led to conclude that the total occupation number operator $\hat{\mathscr{N}}$ of the HP bosons commutes with the spin Hamiltonian. Thus, reducible indexes $\rket{\nu}$ containing the same number of HP bosons form a subspace orthogonal to all others. Since the total spin of the radical is $J_0 = 5/2$, there must be $2J_0+1=6$ orthogonal subspaces $\mathscr{B}_n$, each characterized by the total number of bosons $n \in \{0,1,2,3,4,5\}$ each basis element of the subspace contains. The dimension $\Omega_{\mathpzc{A},n}$ of these subspaces is readily determined, for example, by means of the generating function $G_{\mathpzc{A},\Omega}(q)$, Eq. \eqref{eq:generating_function_G}.  In the case of \ce{^{.}CH2OD}, we have
				\begin{equation}
				\begin{split}
				G_{\mathpzc{A},\Omega}(q) & = [2]^3_q \cdot [3]_q = \sum^5_{n=0} \Omega_{\mathpzc{A},n} \ q^n \\
				& = 1 + 4 q + 7q^2 + 7q^3 + 4 q^4 + q^5 \ . 
				\end{split}
				\end{equation}
So, instead of diagonalizing the matrix $\mathscr{H}_{spin}$, which is of dimension $24$, to obtain its eigenvalues and eigenvectors, we can alternatively accomplish the same by diagonalizing four smaller matrices: two of dimension $4$ and the other two of dimension $7$. The subspace $\mathscr{B}_0$ and $\mathscr{B}_{2J_0}$ are always of dimension one, and correspond to multispin states whereby all the individual spins are in their maximum weight states, \emph{i.e.} $m_i = j_i \ \forall i$ (for $\mathscr{B}_0$) and lowest weight (for $\mathscr{B}_{2J_0}$), respectively. 
\par The basis of the subspaces $\{\mathscr{B}_n\}$ may be determined by writing all the possible strings of irreducible indexes: \emph{i.e.} $\{ \ket{0,0,0,0}, \ket{0,0,0,1}, \ldots, \ket{1,1,1,1}, \ket{1,1,1,2} \}$, and then grouping them according to the total number of HP bosons $n$ they contain. As we have seen above, strings containing the same $n$ HP bosons span the subspace $\mathscr{B}_n$. An even more convenient alternative in determining the basis is to exploit the mapping $\eta_0$, Eq. \eqref{eq:CH2OD_nu}, and the particle-hole symmetry, Eq. \eqref{eq:particle-hole_symm}. For instance, the basis of $\mathscr{B}_1$ are certainly $\{\ket{0,0,0,1},\ket{0,0,1,0},\ket{0,1,0,0},\ket{1,0,0,0}\} \equiv \{\rket{1},\rket{3},\rket{6},\rket{12}\}$. Then, due to the particle-hole symmetry, we know that the complements of $\mathscr{B}_1$'s basis, \emph{i.e.}~$ \{\rket{22},\rket{20},\rket{17},\rket{11}\}$, span $\mathscr{B}_{4}$. The subspaces $\mathscr{B}_n$ for \ce{^{.}CH2OD}, their dimension and basis elements are reported in Table \ref{tab:CH2OD_table}.
\begin{table}[htb]
\centering
\begin{tabular}{|c|c|c|}
\hline 
Subspace $\mathscr{B}_n$, $n(\nu)=\sum^4_{i=1} \nu_i$ & Dimension of subspace & Basis elements, $\rket{\nu}$ \\
\hline \hline
$\mathscr{B}_0$ & 1 &$\rket{0}$\\
$\mathscr{B}_1$ & 4 & $\rket{1}$, $\rket{3}$, $\rket{6}$, $\rket{12}$\\
$\mathscr{B}_2$ & 7 & $\rket{2}$, $\rket{4}$, $\rket{7}$, $\rket{9}$, $\rket{13}$, $\rket{15}$, $\rket{18}$\\
$\mathscr{B}_3$ & 7 & $\rket{5}$, $\rket{8}$, $\rket{10}$, $\rket{14}$, $\rket{16}$, $\rket{19}$, $\rket{21}$\\
$\mathscr{B}_4$ & 4 & $\rket{11}$, $\rket{17}$, $\rket{20}$, $\rket{22}$\\
$\mathscr{B}_5$ & 1 & $\rket{23}$\\
\hline
\end{tabular}
\caption{\textit{Orthogonal subspaces $\mathscr{B}_n$ and their dimensions for the radical \ce{^{.}CH2OD}, assuming a spin Hamiltonian of the generic form given in Eq. \eqref{eq:H_spin scalar-like}. The integer representation $\rket{\nu}$ of the basis elements spanning each subspace depends on the order chosen when constructing the tensor space $\mathcal{H}$, Eq. \eqref{eq:CH2OD_Hilbert_space}}.}
\label{tab:CH2OD_table}
\end{table}
\par Once we know the basis elements of the various subspaces, we can construct their matrix representations through Eq. \eqref{eq:spin_hamil}. To save computational time, we can make use of Eq. \eqref{eq:T-symmetry_matrix_repre}, or equivalently \eqref{eq:T-symmetry_matrix_repre_0}. The eigenvalues and eigenvectors of $\mathscr{H}_{spin}$ then follow from the diagonalization of these smaller matrices. In addition, the density $\zeta$, Eq. \eqref{eq:upper_bound_zeta_HS}, of the radical's spin Hilbert space cannot exceed $0.2291\overline{6}$. In other words, the percentage of $\mathcal{H}$ which effectively encodes the dynamics of the system cannot be greater than $\sim 23\%$ of $\mathcal{H}$. The eigenvectors and eigenvalues of the radical's Liouvillian operator can be easily computed from Eqs. \eqref{eq:supermatrix_U} and \eqref{eq:superdiagonal_D}, respectively.
\section{Conclusion}
We have shown that through the row index compression map $\eta_0$ and the HP transformation, one can easily generate good quantum numbers (in general, integers defined by some composition law) which characterize different subspaces of the Hilbert space. The great merit of this technique is that neither the matrix $\mathscr{H}_{spin}$ nor $\mathfrak{L}_{spin}$ needs to be constructed in order to perform the diagonalization: all one needs are the submatrices $\mathscr{B}_n$ of $\mathscr{H}_{spin}$. But these can also be easily constructed on account of Eqs. \eqref{eq:spin_hamil} and \eqref{eq:total_boson_constraint}. The procedure is as follows: 1) group together all reducible indexes $\ket{\nu}$ containing the same total number of HP bosons $n$, 2) construct the submatrix $\mathscr{B}_n$ for each group using Eq. \eqref{eq:spin_hamil}, and 3) diagonalize the obtained submatrices.
\par Naturally, the presented block diagonalization scheme of the multispin Hamiltonian through the use of $\eta_0$ and the HP bosons reduces exponentially the computational complexity of the problem of finding the eigenvalues and eigenvectors of the latter. We have also shown that the same eigenvectors and eigenvalues can be used to derive the corresponding eigenvectors and eigenvalues of the corresponding Liouvillian superoperator $\sop{L}_{spin}$ of $\op{H}_{spin}$, thus saving us from the even higher computational cost of diagonalizing the former all over again. Furthermore, we emphasize that this diagonalization scheme should be seen as the first step towards a more robust, efficient and analytical procedure to overcome the curse of dimensionality, possibly scaling polynomially with $N$. We mention that once the submatrices have been obtained, one can employ other efficient numerical protocols to diagonalize the larger ones. This will further minimize the computational cost.
\par Besides allowing an easy, yet analytic, monitoring of the various multispin states and transitions (because $\eta_0$, like all $\eta_c$, is bijective) -- not to mention the yoke of carrying along long strings of indexes which it frees us from -- another key advantage of the mapping $\eta_0$ has to do with the fact that it indubitably allows an analytic characterization of multispin Hamiltonians without much effort as we saw in sections \ref{sec:block-diag_spin_Hamil} and \ref{sec:dimension_block_matrices} where we were able to predict how many subspaces we should be expecting and their dimensions. We call the reader's attention to the fact that such results were easily obtained due to the connection the mapping $\eta_0$ and HP bosons create between spin composition and enumerative combinatorics\cite{art:Gyamfi-2018}. 
\par That spin Hamiltonians of the form in Eq. \eqref{eq:H_spin scalar-like} decompose into submatrices is \emph{per se} not any news. There are a number of ways to prove the same. Corio, for example, has extensively treated the topic in his seminal book\citep{book:Corio-1966}. What the literature lacked, as far as we know, is a general, comprehensive and systematic way to effectively realize such decomposition. And that is what the present article aims at.    
\par The procedure we have presented can be easily extended to anisotropic multispin Hamiltonians. There, the inherent symmetries of the coupling tensors together with other symmetries may be exploited in determining the good quantum numbers in the integer representation. Moreover, the method presented also lends itself to the study of various quantum systems defined on finite Hilbert spaces, these include a variety of spin lattice models and many variants of the simple symmetric exclusion processes (see for example \cite{art:Mendonca-2013}), to say the least.
\par Finally, we remark that the presence of square roots of operators in the HP transformation may prove to be inconvenient in certain scenarios. By resorting to the Schwinger bosons\citep{book:Auerbach-1998}, one avoids these square roots at the cost of dealing with two flavors. The use of Schwinger bosons is worth pursuing as it might probably lead to an even more efficient diagonalization scheme.

\section*{Acknowledgment}
The authors are very grateful to Prof. Antonino Polimeno (Univ. of Padova) and Prof. Giorgio J. Moro (Univ. of Padova) for their discerning comments upon reading an abridged version of the manuscript. JAG would also like to thank Mr. Andrea Piserchia (SNS) for the many interesting discussions they have had over the past months. The authors employed computer facilities at SMART@SNS laboratory, to which they are greatly grateful. The research leading to these results has received funding from the European Research Council under the European Union's Seventh Framework Program (FP/2007-2013) / ERC Grant Agreement n. [320951].
 
%\section*{References}
\bibliography{Taming_Curse_biblio}
\end{document}